\documentclass[aps,groupedaddress,showpacs,twocolumn]{revtex4}

\usepackage{epsfig}
\usepackage{psfig}
\begin{document}

\title{Modulational instability, solitons and beam propagation in
spatially nonlocal nonlinear media}

\author{W. Kr\'olikowski$^{1}$, O. Bang$^{2,3}$, N.I.
Nikolov$^{3,4}$,   D. Neshev$^{5}$, J. Wyller$^{1,5,6}$, J.J.
Rasmussen$^4$, and   D. Edmundson$^7$}

\address{$^1$ Laser Physics Centre, Research School of Physical Sciences
            and Engineering, Australian National University,
            Canberra ACT 0200, Australia}

\address{$^2$ Research Center COM, Technical University of Denmark,
            DK-2800 Kongens Lyngby, Denmark}

\address{$^3$ Informatics and Mathematical Modelling, Technical
University of  Denmark, DK-2800 Kongens Lyngby, Denmark}

\address{$^4$ Ris{\o} National Laboratory, Optics and Plasma Research Department, OPL-128, P.O. Box. 49, DK-4000 Roskilde, Denmark}

\address{$^5$ Nonlinear Physics Group, Research School of Physical
            Sciences and Engineering, Australian National University,
            Canberra ACT 0200, Australia}

\address{$^6$ Department of Mathematical Sciences and Technology,
            Agricultural University of Norway, P.O. Box 5003,
            N-1432 {\AA}s, Norway}

\address{$^7$ ANU Supercomputer Facility, Australian National University,
            Canberra ACT 0200, Australia}

\begin{abstract}
We present an overview of recent advances in the understanding
of optical beams in nonlinear media with a spatially nonlocal nonlinear 
response.
We discuss the impact of nonlocality on the modulational instability
of plane waves, the collapse of finite-size beams, and the formation and
interaction of spatial solitons.
\end{abstract}



\maketitle


\section{Introduction}
A soliton is a localised wave that propagates without change through a
nonlinear medium. Such a localized wave forms when the dispersion or
diffraction associated with the finite size of the wave is balanced by
the nonlinear change of the properties of the medium induced by the
wave itself.
Solitons are universal in nature and have been identified in such physical
systems, as fluids, plasmas, solids, matter waves, and classical
field theory.
Spatial optical solitons - self-trapped light beams - have been
proposed as building blocks in future ultra-fast all-optical devices.
Spatial solitons can be used to create reconfigurable optical circuits
that guide other light signals.
Circuits with complex functionality and all-optical switching or
processing
can then be achieved through the evolution and interaction of one or more
solitons \cite{SnyderLadouc}. This concept has now been verified in
several
optical materials \cite{Yuri_book,science} and a number of new soliton
effects have
emerged through these studies, such as fusion, fission, and formation
of bound states. Due to the development of materials with stronger
nonlinearities the optical power needed to create such virtual circuits
has been reduced to the milliwatt and even microwatt level, bringing
the concept nearer to practical implementation.

The soliton concept is an integral part of studies of the coherent
excitations of Bose-Einstein condensates (BECs) \cite{BEC1}. Such BECs,
inherently have a spatially nonlocal nonlinear response due to the finite
range of the inter-particle interaction potential. Spatial nonlocality,
which is already an established concept in plasma physics
\cite{Litvak75,juul,df}, means that the response of
the medium at a particular point is not determined solely by the wave
intensity at that point (as in local media), but also depends on the
wave intensity in its vicinity.
The nonlocal nature often results from a transport process, such as
atom diffusion \cite{suter}, heat transfer \cite{thermal1,thermal2} or 
drift of electric charges \cite{GatzHerman}. It can also be induced by  a long-range molecular interaction as in nematic liquid crystals which exhibit orientational nonlocal nonlinearity \cite{lc1,Assanto}.
Nonlocality is thus a generic feature of a large number of nonlinear
systems.  It has also recently become important in
optics \cite{Snyder97,Snyder99,Bertolotti}.
Although nonlocality can have a considerable impact on many nonlinear
phenomena, studies of nonlocal nonlinear effects are still in their
infancy.

In this paper we will discuss the role of spatial nonlocality in phenomena
associated with propagation of optical beams in nonlinear media.
These include modulational instability of plane waves, structural
stability of localized beams and interaction of solitons.

\section{Model}

We consider an optical beam propagating along the $z$-axis of a
nonlinear medium with the scalar electric field
$E(\vec{r},z)=\psi(\vec{r},z)\exp(iKz-i\Omega t)+c.c.$
Here $\vec{r}$ spans a D-dimensional transverse coordinate
space, $K$ is the wavenumber, $\Omega$ is the optical frequency, and
$\psi(\vec{r},z)$ is the slowly varying amplitude.
We assume that the refractive index change $N(I)$ induced by the beam
with intensity $I(\vec{r},z)=|\psi(\vec{r},z)|^2$ can be described
by the phenomenological nonlocal model
\begin{equation}
      \label{nonlocal}
      N(I)= s\int R(\vec{\xi}- \vec{r})I(\vec{\xi},z)d\vec{\xi},
\end{equation}
where the integral $\int d\vec{r}$ is over all transverse dimensions and
$s=1$ ($s=-1$) corresponds to a focusing (defocusing) nonlinearity.
The response function $R(\vec{r})$, which is assumed to be real, localized
and symmetric (i.e. $R(\vec{r})=R(r)$, where $r=|\vec{r}|$), satisfies the
normalization condition $\int R(\vec{r})d\vec{r}=1$.
This model of nonlinearity leads to the following nonlocal nonlinear
Schr\"{o}dinger (NLS) equation governing  the evolution of the beam
\begin{equation}
i\partial_z\psi +
\frac{1}{2}\nabla_{\perp}^2\psi + N(I)\psi = 0.
\label{nonlocalNLS}
\end{equation}
%
\begin{figure}
      \centerline{\psfig{file=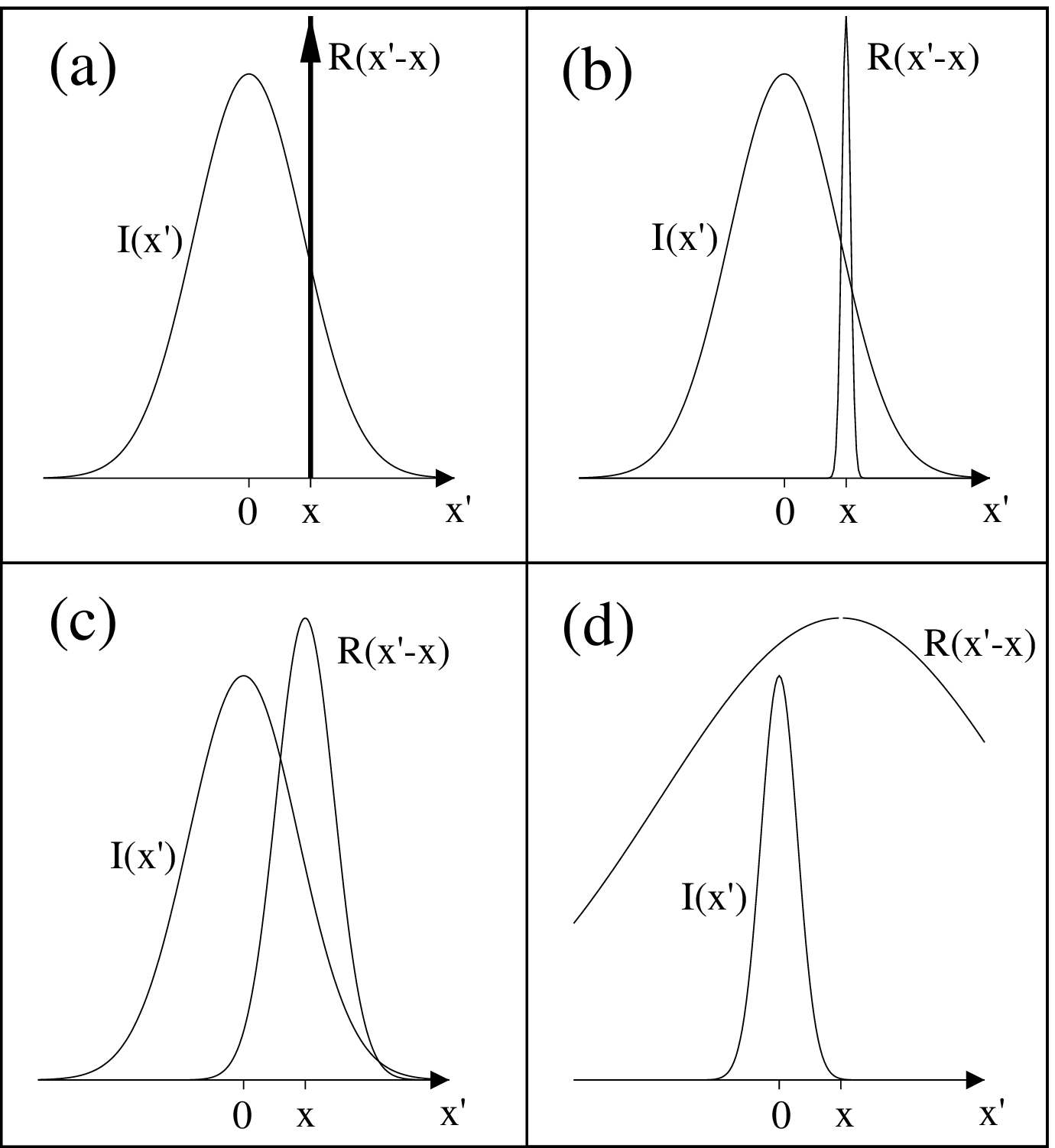,width=0.5\linewidth}}
      \caption{Various degrees of nonlocality, as given by the width of
the
nonlocal response function $R(x)$ and the intensity profile of the beam
$I(x)$. Shown are (a) local, (b) weakly nonlocal, (c) general (nonlocal) and
(d) highly  nonlocal responses.  \label{degrees}}
\end{figure}
 
The width of the response function $R(r)$ determines the degree of
nonlocality.
For a singular response, $R(r)=\delta(r)$ [see Fig.\ref{degrees}(a)], the refractive index change becomes a local function of the light intensity, $N(I)=
sI(\vec{r},z)$, i.e., the refractive index change at a given point is
solely determined by the light intensity at that very point, and
Eq.(\ref{nonlocalNLS}) simplifies to the standard NLS equation
\begin{equation}
   i\partial_z\psi + \frac{1}{2}\nabla_{\perp}^2\psi + s\psi|\psi|^2 = 0.
   \label{NLS}
\end{equation}
With increasing width of $R(r)$ the light intensity in the
vicinity of the point $\vec{r}$ also contributes to the index change at
that point [see Fig.\ref{degrees}].
While Eq.~(\ref{nonlocal}) is a phenomenological model, it nevertheless
adequately describes several physical situations where the nonlocal nonlinear
response is due to  various transport effects, such as heat conduction or
diffusion of molecules or atoms.

Before proceeding further it is worthwhile to mention two important 
physical situations where the convolution term in Eq.~(\ref{nonlocalNLS}) 
can be represented in a simplified form, which allows for an extensive 
analytical treatment of the resulting equation. 
First is the so-called weak nonlocality limit when
the width of the response function is much less than the spatial extent
of the beam [see Fig.\ref{degrees}(b)].
In this case one can formally expand $I(\vec{r})$ in a Taylor series and 
retain only the first significant  terms, which gives the 
following simple form of the nonlinearity
\begin{equation}
     \label{Nweak}
     N(I) = s(I + \gamma\nabla_{\perp}^2I), \quad
     \gamma = \frac{1}{2D}\int r^2 R(r) d\vec{r},
\end{equation}
where $\gamma$ is a measure of the strength of the nonlocality.
Here the nonlocal contribution to the Kerr-type local nonlinearity is
reflected by the presence of the Laplacian of the wave intensity.
A nonlinearity in this particular form appears in fact 
naturally in the theory  of nonlinear effects in plasma \cite{df}.
It has been
shown recently that the one-dimensional version of Eq.~(\ref{nonlocalNLS})
with nonlinearity (\ref{Nweak}) supports propagation of stable bright and
dark solitons \cite{wkob}.
Another limiting case, the so-called highly nonlocal limit, refers to
the situation when the nonlocal response function is much wider than the
beam itself [see Fig.\ref{degrees}(d)]. In this case 
\begin{equation}
   N(I)=sR(r)P, \quad  P=\int Id\vec{r},  
\end{equation}
where $P$ is the total power of the beam.
Interestingly, in this case the propagation equation becomes local
and linear. It describes the evolution of an optical beam trapped in an
effective waveguide structure with the profile given by the
nonlocal response function. This highly nonlocal limit was first
explored  by  Snyder and Mitchell in the context of the 
"accessible solitons" \cite{Snyder97}.

Although it is quite apparent in  several physical situations that
the nonlinear response in general is nonlocal (as in the case of thermal
lensing), the nonlocal contribution to the refractive index change is
often neglected \cite{dark1,dark2}.
This is justified if the spatial scale of the beam remains large compared
to the characteristic response length of the medium (given by the
width of the response function). However, for very narrow beams or beams
with fine spatial features (such as dark and bright solitons) the
nonlocality can be of crucial importance and should  be taken into account
in the ensuing theoretical model.

\section{Modulational instability}

Modulational instability (MI) constitutes one of the most
fundamental effects associated with wave propagation in nonlinear
media. It signifies the exponential growth of a weak perturbation
of the wave  as it propagates. The gain leads to amplification of sidebands, 
which break up the otherwise uniform wave and generate fine 
localized structures (filamentation).
Thus, it may act as a precursor for the formation of bright
solitons. Conversely, the existence of stable dark 
solitons requires the absence of MI of the constant intensity
background. The phenomenon of MI has been identified and studied in
various physical systems, such as fluids \cite{mi-fluid}, plasma
\cite{mi-plasma}, nonlinear optics \cite{mi-nlo1},  discrete
nonlinear systems \cite{discrete-mi-molchain}, 
waveguide arrays, and Fermi-resonant interfaces
\cite{miller-mi}.
It has been shown that MI is strongly affected by mechanisms 
such as saturation of the nonlinearity
\cite{mi-saturation}, coherence properties of optical beams 
\cite{mi-coherence}, and linear and nonlinear 
gratings \cite{corney-mi}.  
The model (\ref{nonlocalNLS}) permits plane
wave solutions of the form
\begin{equation}
  \psi(\vec{r},z) = \sqrt{\rho_0}\exp(i\vec{k}_0\cdot\vec{r}-i\beta z),
  \label{planewave}
\end{equation}
where $\rho_0>0$ is the wave intensity. The parameters satisfy the
same dispersion relation $\beta=k_0^2/2-s\rho_0$ as for the
standard local NLS equation, because the response function is
normalized. We now consider perturbed plane wave solutions in the
form
\begin{eqnarray}
   \label{wavepert}
   \psi(\vec{r},z)= [\sqrt{\rho_0}+a_{1}(\vec{\xi},z)]
   \exp(i\vec{k}_0\cdot\vec{r}-i\beta z),
\end{eqnarray}
where
\begin{equation}
   a_1(\vec{\xi},z)=\int\widetilde{a}_1(\vec{k})\exp(i\vec{k}\cdot\vec{\xi}+
   \lambda z)d\vec{k},\quad \vec{\xi}=\vec{r}-\vec{k}_{0}z,
\end{equation}
 is the complex amplitude of the small perturbation referred to a coordinate 
frame moving with the  group velocity $\vec{k}_0$.
One can show that $\lambda$ satisfies the equation \cite{MI-wk}
\begin{equation}
      \lambda^2 = -k^2\rho_0\left[\frac{k^2}{4\rho_0}-s\widetilde{R}(k)
\right],
      \label{generaleigenvalue}
\end{equation}
where  $ k=|\vec{k}|$ denotes the spatial frequency, 
$\widetilde{R}(k)$
is the Fourier spectrum of $R(r)$. The sign of $\lambda^2$ determines the 
stability of the solution. For $\lambda^2>0$ the perturbation grows 
exponentially during propagation with the growth rate or gain
given by Re$\{\lambda\}$, indicating MI. Equation
(\ref{generaleigenvalue}) shows that the stability properties of
the plane wave solutions are completely determined by the spectral
properties of the nonlocal response function. A detailed analysis of
all the possible scenarios is given in \cite{Wyller}. Here we
summarize the main points.

\begin{figure}
  \centerline{\psfig{file=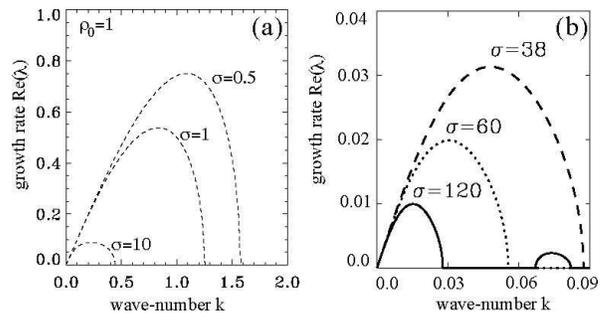,width=0.9\linewidth}}
  \caption{MI gain profile in self-focusing nonlocal media with
  Gaussian response function for $\rho_0$=1 (a) and with 
  generalized Lorentzian response function for 
  $n$=2 and $\rho_0$=1.25 (b). \label{mi_1}}
\end{figure}

The Fourier spectrum of typical
response functions, such as Gaussians, Lorentzians, and
exponentials, is positive definite. Therefore, for a defocusing
nonlinearity ($s$=$-1$), 
$\lambda^2<0$ and the plane wave solutions 
are always stable.
In a focusing medium ($s$=$+1$) we
always have $\lambda^2>0$, in a certain wavenumber band
symmetrically centered about the origin, where $k$ is sufficiently
small. This means that the system will always exhibit a long
wave MI in the focusing case, independent of the precise details 
of the response function.  Nonlocality tends to suppress the
instability, decreasing the growth rate and the width of the 
instability band. However, nonlocality can never eliminate the 
instability completely. This effect is illustrated in   
Fig.~\ref{mi_1}(a), where we show the instability growth rate in 
the case of a one-dimensional Gaussian response 
function $R(x)=\exp(-x^2/\sigma^2)/(\sqrt{\pi}\sigma)$.

\begin{figure}
     \centerline{\psfig{file=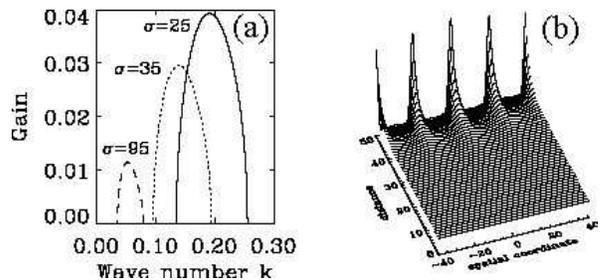,width=0.9\linewidth}}
     \caption{MI in self-defocusing nonlocal media
     with response function (\ref{response}).
     (a) MI gain for $n$=2 and $\rho_0$=1.
     (b) Development of instability for $n$=6 and $\rho_0$=3.
     \label{mi_2}}
\end{figure}

A drastically different behavior is observed for  response
functions whose spectrum is not sign definite. Let us consider the
generalized Lorentzian \cite{Lisak}
\begin{equation}
   \label{response}
   R(x) = \frac{\pi}{n\sigma}\frac{1}{\sin(\frac{\pi}{2n})}\frac{1}
   {1+(\frac{x}{\sigma})^{2n}}, \quad n=2,3,...
\end{equation}
which for large $n$ approximates the rectangular function.
The spectrum of this function, which for $n$=2 is
$\tilde{R}(k)=\exp(-\sigma |k|/\sqrt{2})\left [\cos(\sigma
|k|/\sqrt{2})+\sin(\sigma |k|/\sqrt{2})\right ]$ exhibits periodic
changes of its sign. For focusing nonlinearity this may  lead
to the appearance of higher order instability bands in sufficiently nonlocal media [see Fig.~\ref{mi_1}(b) for $\sigma$=120]. Even more dramatic is the behaviour of
the plane wave solution in a defocusing medium, where this type of
nonlocality may promote a high frequency instability of the otherwise
stable wave [see Fig.~\ref{mi_2}(a)]. In Fig.~\ref{mi_2}(b) we
show the development of the instability of the plane wave. The
response function (\ref{response}) has recently been discussed in
the context of MI of partially coherent waves, where it represents 
the degree of coherence of the plane wave. 
Interestingly, it also promoted MI in that case
\cite{Lisak}. It is worth mentioning that the spectrum of
(\ref{response}) for any value of the exponent $n>2$ contains an
oscillatory factor  responsible for the formation of higher order
gain bands in a way analogous to the case $n=2$.

The above-mentioned stability properties of the plane wave may seem
surprising in view of the fact that the nonlocality
acts to smooth out any sharp modulations of the wavefront.
This is a generic property of the nonlocality independent of the
particular functional representation. One would therefore naively
expect identical stability properties for all physically
reasonable response functions. However, one can also look at the
action of the nonlocality from a different perspective. In the
Fourier domain the nonlocality acts as a filter  with variable
transmission determined by the form of the spectrum of the
response function. For many nonlocal models, such as that with a
Gaussian response function, the transmission characteristic of the filter has
the form of a well-behaving, sign-definite function. However, in
cases, such as those modelled  by (\ref{response}), this filter
not only modulates the amplitude of the spectral components of the
signal beam (perturbation to the plane wave), but also inverts the
phase of selected components (bands). As this inversion is equivalent
to a change of the sign of the nonlinearity (say, from defocusing
to focusing), it leads, for instance, to  amplification of certain
harmonics in a defocusing nonlinear media. Recently Peccianti
{\em et al.} conducted the first experimental studies of MI in
a nonlocal medium \cite{Assanto_MI}. They used nematic liquid
crystals, which are known to exhibit orientational nonlocal
nonlinearity with a sign-definite exponential response. The effect
of suppression of MI was clearly demonstrated.

\section{Beam collapse}

Collapse is a phenomenon well-known in the theory of wave
propagation in nonlinear focusing media. It refers 
to the situation when strong self-focusing of a beam leads to a catastrophic
increase (blow-up) of its intensity in a finite time or after a finite
propagation distance \cite{RasRyp86,Berge98,KivPel00}.
Collapse has been observed in plasma waves \cite{Wong}, 
electromagnetic waves or laser beams \cite{Garmire}, BEC's 
\cite{Hulet99}, and even capillary-gravity waves on deep water 
\cite{water}.

Strictly speaking, the existence of collapse is an artefact of the 
model equation and  signals the limit of its  applicability.
In the vicinity of the collapse regime some additional (unaccounted 
for) physical processes will come into play  and stop the blow-up. 
Nevertheless, ``collapse-like'' (or quasi-collapse) dynamics can
still occur in real physical systems when nonlinearity leads to strong
energy localisation. In fact, recent experiments with ultra cold
gases provided clear signatures of the collapse-like dynamics
of atomic condensates \cite{Hulet99,Hulet00}.

In the past there have been a few attempts to determine the role of nonlocality in
development of the collapse of finite-size beams. Turitsyn was the first to prove analytically the arrest of
collapse for a specific choice of the nonlocal nonlinear response
\cite{Tur85}. Recently Perez-Garcia {\em et al.} discussed   the collapse suppression  in the case of weak nonlocality, in a direct application for a BEC \cite{PerKonGar02}. The  analysis of the collapse conditions in  case of general BEC nonlocal model is difficult and has so far
only been  done  numerically \cite{PerKonGar00}.

Here we present an analytical approach to beam collapse in nonlocal 
media, which is based on the technique introduced in \cite{Bang02},
but which extends and generalizes the results to be valid for much 
more general response functions. We consider the general case of {\em
symmetric, but otherwise arbitrarily shaped, non-singular response 
functions} and prove rigorously that a collapse cannot occur.

For localized or periodic solutions Eqs.~(\ref{nonlocal}) and
(\ref{nonlocalNLS})
conserve the power (in optics) or number of atoms (for BEC) $P$ and the
Hamiltonian $H$,
\begin{equation}
\label{power-hamiltonian}
    P = \int I d\vec{r}, \quad
    H = \frac{1}{2}||\nabla \psi||_2^2-\frac{1}{2}\int NId\vec{r},
\end{equation}
where $||\psi||_2^2\equiv\int|\psi|^2d\vec{r}$.
In the local limit when the response function is a delta-function,
the nonlinear response has the form $ N(I) = sI$, as in local optical Kerr
media described by the conventional NLS equation and in BEC's
described by the standard Gross-Pitaevskii equation. In this local
limit multidimensional beams with a power higher than a certain
critical value will experience unbounded self-focusing and {\em
collapse} after a finite propagation
distance. 

It can be easily shown that in the two extreme limits of a weakly
and highly nonlocal nonlinear response the collapse is prevented
\cite{df,ParSalRea98}. Here we consider the general case of {\em
symmetric, but otherwise arbitrarily shaped, non-singular response 
functions}.
Introducing the D-dimensional Fourier transform and its inverse, it is straightforward to show that for
$N(I)$ given by Eq.~(\ref{nonlocal}), when $s$=1 the following 
relations hold \cite{Bang02}
\begin{eqnarray}
   |\tilde{I}(\vec{k})|& =& |\int I(\vec{r}){\rm e}^{i\vec{k}\cdot\vec{r}}
   d\vec{r}| \le P, \nonumber \\
   \int NId\vec{r} & = &  \frac{1}{(2\pi)^D} \int \widetilde{R}(\vec{k}\,)
   |\tilde{I}(\vec{k}\,)|^2d\vec{k}. 
\end{eqnarray}
For any response functions for which the spectrum
$\widetilde{R}(k)$ is absolute integrable, we then have
\begin{equation}
   \label{Ibound}
   |\int NId\vec{r}| \le P^2R_0, \quad
   R_0\equiv \frac{1}{(2\pi)^D} \int|\widetilde{R}(k)|d\vec{k},
   \nonumber
\end{equation}
and hence we get $|H|\ge||\nabla \psi||_2^2/2-R_0P^2/2$. 
This inequality shows that the gradient norm $||\nabla \psi||_2^2$ is
bounded from above by the conserved quantity $2|H|+R_0P^2$
for all symmetric response functions under the only requirement that 
their spectrum is absolute integrable.
It represents a rigorous proof that a collapse with the
wave-amplitude locally going to infinity cannot occur in BEC's or
Kerr media with a nonlocal nonlinear response for any
physically reasonable response functions, such as Gaussians, 
exponentials, sech-shaped, and even generalized Lorentzian
response functions of the type (\ref{response}).

This result represents a considerable generalization of the proof presented 
in \cite{Bang02}, which required the spectrum of the response function 
to be {\em strictly positive definite}.

\begin{figure}
  \centerline{\psfig{file=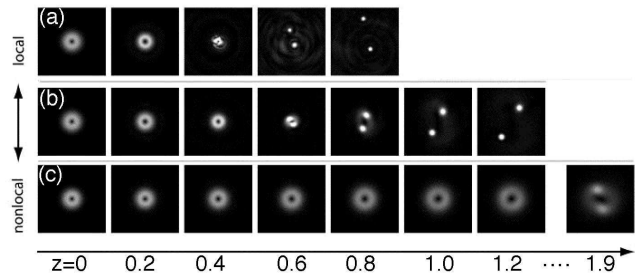,width=1.0\linewidth}}
  \caption{Propagation of a vortex beam (charge 1) in a
  self-focusing nonlocal Kerr-like medium. The nonlocality parameter
 $\sigma=0$ (a), $\sigma=1$ (b), $\sigma=10$, (c).
  \label{vortex}}
\end{figure}

We emphasize that although the nonlocality prevents collapse, it does
 not remove the collapse-like dynamics of high-power beams. To the
contrary, as long as the beam width is much larger than the extent of
the nonlocality (width of the response function), the beam will
naturally contract as in the local case.  However the collapse will be
arrested when the width becomes comparable with that of the response
function. Finally, during the stationary phase of the beam propagation,
its width will be comparable with the extent of the nonlocality.

The stabilising effect of the nonlocality can be further
illustrated by considering the propagation of a vortex beam in a
self-focusing medium. Such a beam is characterized by a  bright ring with
a helical  phase front with so-called
charge, defined as a closed loop contour integral of the wave
phase modulo $2\pi$. A typical example is the Gaussian-Laguerre beam
\begin{equation}
   \label{vortexbeam} 
   \psi(\vec{r}) = r\exp\left[-(r/r_0)^2\right]\exp(i\phi),
\end{equation}
where $r$ and $\phi$ are the radial and angular coordinates,
respectively.  This beam represents  a  vortex of charge one.
Beams of such structure have been considered as candidates for
vortex-type solitons in nonlinear self-focusing media
\cite{Yang03}. However, it is well known that  vortex  beams
cannot form stable stationary structures  and disintegrate rather
quickly when launched in self-focusing local Kerr 
nonlinear media \cite{brakeup}.
One can expect to achieve a stabilization of the vortex
beam by utilizing the nonlocal character  of the nonlinearity. If
the extent of the nonlocality is comparable with the size of the
vortex beam then the resulting  refractive index change  will  
form of a broad circular waveguide, which could trap the vortex
beam ensuring its stable propagation. In Fig.~(\ref{vortex}) we
illustrate the effect of the nonlocality on propagation of a
vortex beam in the form of Eq.~(\ref{vortexbeam}). We numerically
integrated Eq.~(\ref{nonlocalNLS}) assuming a Gaussian nonlocal 
response  $R(x)=\exp(-x^2/\sigma^2)/(\sqrt{\pi}\sigma)$.  The plots in each row show the light
intensity distribution at various propagation distances.
The upper row represents local Kerr medium ($\sigma$=0). It
is clearly seen that the vortex beam experiences strong focusing
and quickly breaks up into two fragments. As the nonlocality engages 
($\sigma=1$, middle row) the vortex instability is restrained 
and break-up occurs only after longer propagation. The distance of
stable propagation significantly increases 
when  nonlocality becomes larger ($\sigma=10$, bottom row).

\section{Nonlocal structure of parametric solitons}

\noindent
Unlike Kerr solitons, the formation of solitons in quadratic nonlinear (or
$\chi^{(2)}$) materials does not involve a change of the refractive index
\cite{Bur02}.
Thus the underlying physics of quadratic solitons is often obscured by
the
mathematical model. Only recently Assanto and Stegeman used the cascading
phase shift and parametric gain to give an intuitive interpretation of
self-focusing, defocusing, and soliton formation in $\chi^{(2)}$
materials
\cite{Ass02}.
Nevertheless several features of quadratic solitons are still without a
physical interpretation, such as certain structural properties
and the formation of bound states.
We will show below that quadratic solitons can be described by 
nonlocal  models. Such models provide simple physical explanations of
these properties and many more, building on a simple waveguide
analogy   \cite{Sha02,Con03,Nik03}.

Consider a fundamental wave (FW) and its second harmonic (SH) propagating
along the $z$-direction in a $\chi^{(2)}$ crystal under conditions for
type
I phase-matching. The normalized dynamical equations for the slowly
varying
envelopes $E_{1,2}(x,z)$ are then \cite{Men94Ban97}
\begin{eqnarray}
      \label{chi2dynam}
      & & i\partial_zE_1 + d_1\partial_x^2E_1 + E_1^*E_2 \exp(-i\beta z) = 0,
\nonumber \\ 
     & &  i\partial_zE_2 + d_2\partial_x^2E_2 + E_1^2 \exp(i\beta z) = 0.
\end{eqnarray}
In the spatial domain $d_1\approx 2d_2$, $d_{1,2}>0$, and $x$ represents
a
transverse spatial direction.
In the temporal domain $d_{1,2}$ is arbitrary and $x$ represents time.
$\beta$ is the normalized phase-mismatch.
Physical insight into Eqs.~(\ref{chi2dynam})
may be obtained from the cascading limit, in which the phase-mismatch is
large, $\beta^{-1}\rightarrow0$.
Writing $E_2$=$e_2\exp(i\beta z)$ and assuming slow variation of
$e_2(x,z)$ gives the NLS equation
$i\partial_z E_1 + d_1\partial_x^2E_1 + \beta^{-1}|E_1|^2E_1 = 0$,
with $e_2$=$E_1^2/\beta$.
However, this model wrongly predicts several features that are known not
to exist in Eqs.~(\ref{chi2dynam}) and even for stationary solutions it
is
inaccurate, since the term $\partial_x^2E_2$ is neglected \cite{Nik03}.
To obtain a more accurate model we assume a slow variation of the SH
field
$e_2(x,z)$ in the propagation direction only (i.e., only $\partial_z e_2$
is neglected), which leads to the {\em nonlocal equation for the FW}
\begin{eqnarray}
   \label{nonlocal_dynam}
   & & i\partial_z E_1 + d_1\partial_x^2E_1 + \beta^{-1} N(E_1^2)E_1^* 
       = 0, \nonumber \\
   & & N(E_1^2) = \int_{-\infty}^\infty R(x-\xi)E_1^2(\xi,z) d\xi,
\end{eqnarray}
with $E_2$=$\beta^{-1}N\exp(i\beta z)$. Equations
(\ref{nonlocal_dynam}) show that the interaction between the FW
and SH is equivalent to the FW propagating in a medium with a
nonlocal nonlinearity. In the Fourier domain the response function 
is a Lorentzian
$\widetilde{R}(k)$=$1/(1+s\sigma^2k^2)$, where
$\sigma$=$|d_2/\beta|^{1/2}$ represents the degree of nonlocality
and $s$=sign$(d_2\beta)$. Both Eqs.~(\ref{chi2dynam}) and
(\ref{nonlocal_dynam}) are trivially extended to more transverse
dimensions.

For $s$=+1, where the $\chi^{(2)}$-system (\ref{chi2dynam}) has a
family of bright (for $d_1$$>$$0$) and dark (for $d_1$$<$$0$)
soliton solutions \cite{Bur95}, $\widetilde{R}(k)$ is positive
definite and localized, giving $R(x) =
(2\sigma)^{-1}\exp(-|x|/\sigma)$. It is possible to show, e.g.,
that the nonlocal model (\ref{nonlocal_dynam}) does
not allow collapse in any physical dimension \cite{Nik03}, a known
property of the $\chi^{(2)}$ system (\ref{chi2dynam}) not captured
by the cascading limit NLS equation. The cascading limit
$\beta^{-1}\rightarrow 0$ is now seen to correspond to  the local
limit $\sigma\rightarrow 0$, in which the response function
becomes a delta function, $R(x)\rightarrow\delta(x)$. With the
nonlocal analogy one can further assign simple physically
intuitive models to the weakly nonlocal limit $\sigma$$\ll$1
and the strongly nonlocal limit $\sigma$$\gg$1.

For $s$=$-1$ $\widetilde{R}(k)$ has poles on the real axis and the
response function becomes oscillatory with the Cauchy principal
value $R(x)$=$(2\sigma)^{-1}\sin(|x|/\sigma)$. In this case the
propagation of solitons has a close analogy with the evolution of
a particle in a nonlinear oscillatory potential. In fact, it is
possible to show that the oscillatory response function explains
the fact that dark and bright quadratic solitons radiate linear
waves \cite{Bur95}. 

Equations (\ref{nonlocal_dynam}) show the
important novel result, that in contrast to the conventional
nonlocal NLS equation treated in detail in this work, the nonlocal
response of the $\chi^{(2)}$ system depends on the square of the
FW, not its intensity. Thus, the phase of the FW enters into the
picture and one cannot directly transfer the known dynamical
properties of plane waves and solitons, such as stability. The
general model (\ref{nonlocal_dynam}) and its weakly and strongly
nonlocal limits thus represents novel equations, whose properties
potentially allow to understand yet unexplained dynamical
properties of quadratic solitons. 

In contrast, the stationary properties of nonlocal solitons, such as
the dependence of their profiles  on material parameters, directly apply to
quadratic solitons.
One can show that the structure of the nonlocal equation
governing the stationary fields resulting from (\ref{nonlocal_dynam})
is  identical to the conventional nonlocal model for stationary solitons
and thus it has the same weakly and strongly nonlocal limits with the
same exact bright and dark soliton solutions.
We recently showed that the nonlocal model elegantly explains the
structural properties of both bright and dark solitons and their bound
states and that  it  provides good  approximate quadratic soliton
solutions in large regimes of the parameter space \cite{Nik03}.

\section{Interaction of dark nonlocal solitons }

To this point, we have concentrated on the properties of individual 
beams in nonlocal nonlinear media. Because of the specific nature
of the nonlocality, which results in spatial advancing of the
nonlinearity far beyond the actual spatial location of the beam,
it is natural to expect strong influence of the nonlocality on 
interaction of well separated localised waves and solitons.
For instance, in case of two nearby optical beams, each
of them will induce a refractive index change extending into the
region of the other one, thereby affecting its trajectory. One can
show that in a self-focusing medium nonlocality always provides an 
attractive force between interacting bright solitons.
This effect has been recently demonstrated for the interaction of 
bright solitons formed in a liquid crystal \cite{lc2}. 
It has been shown that even out-of-phase bright spatial solitons, which in a
local medium always repel, experienced strong attraction that can 
only be overcome by a sufficiently large initial divergence of 
the soliton trajectories \cite{lc2}.
As a consequence of the nonlocality induced attraction, bound states
of out-of-phase bright solitons can be formed 
\cite{Halterman}. In this section we will describe a novel
phenomenon of attraction of dark solitons \cite{Kivshar98}
in nonlocal nonlinear media with a self-defocusing
nonlinearity \cite{Nikola04}.

We will concentrate on the interaction of one-dimensional solitons. 
Without loss of generality, we consider the exponential response 
function $R(x)=(2\sigma)^{-1}\exp(-|x|/\sigma)$. Following the
discussion in the preceeding section about the equivalence between 
stationary nonlocal and parametric solitons, and the results of Refs.
\cite{Nik03,Bur95}, the nonlocal NLS equation with such a response
function predicts the existence of single fundamental dark soliton 
solutions with nonmonotonic tails above a certain critical value of 
$\sigma$. As a result bound states involving two or more solitons 
can be formed.

\begin{figure}
   \centerline{\psfig{file=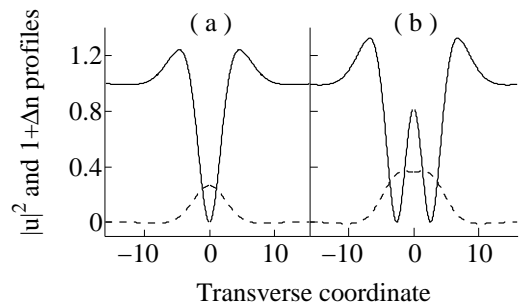,width=0.8\linewidth}}
   \caption{Intensity profile of numerically found stationary 
   dark nonlocal solitons (solid lines). (a) Single dark soliton
   and (b) a bound state of two dark solitons for $\lambda=1$ and $\sigma=4$. Dashed lines indicate the soliton-induced refractive index change. 
   \label{profiles}}
\end{figure}

In Fig.~\ref{profiles} we show examples of numerically found dark
soliton solutions and their bound states.
Dashed line illustrates the soliton-induced waveguide
structure  which guides the soliton. 
These solitons appear to be very robust.

\begin{figure}
      \centerline{\psfig{file=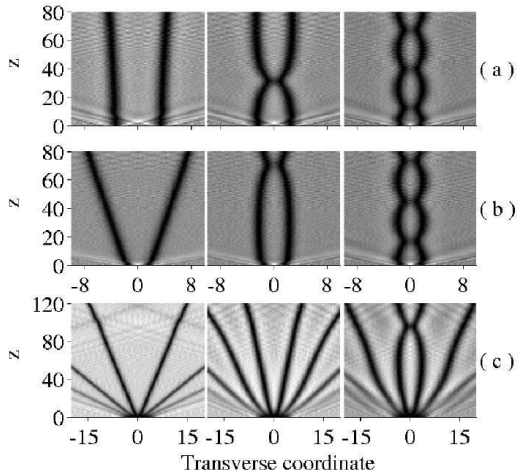,width=0.8\linewidth}}
      \caption{Attraction  of dark nonlocal solitons formed either by
      two closely spaced phase jumps
(a,b) or a dark notch (gap)  in  the initial cw background intensity (c). In (a) the
phase jump is $\pi$ and the degree of nonlocality is $\sigma$=2,
while the initial soliton separation is $x_0$=5.5, 4, 2.5    from
left to right. In (b) the phase jump is $0.95\pi$ and $x_0$=2.5,
while $\sigma$=0.1, 1, 2. In (c) the width of the intensity gap is
7.5, while $\sigma$=0.1, 3, 6. \label{formation}}
\end{figure}

The ability for dark solitons to form bound states and their
subsequent stability is a direct consequence of the {\em
nonlocality-induced long range attraction of solitons}. This
effect can be qualitatively explained using the self-guiding
concept. In a local defocusing medium the refractive index
change corresponding to two distant dark solitons has the form of
two waveguides separated by a region of lower refractive index
(a potential barrier). In the presence of nonlocality  the effect
of the convolution term in Eq.~(\ref{nonlocalNLS}) is to decrease
the index difference between these two separate waveguides  (lower
the barrier)  thereby allowing light to penetrate the area between
solitons. This, consequently, manifests  as soliton attraction. The
attraction of solitons can be clearly demonstrated by simulating
numerically the interaction dynamics of two nearby
solitons. The results of these simulations are summarized in
Fig.~\ref{formation}(a-c). They clearly demonstrate that as the
nonlocality parameter $\sigma$ becomes comparable with the
separation of the solitons they strongly attract and trap each
other and subsequently  propagate together as a bound  state
exhibiting transverse oscillations.

\section{Conclusions}

In conclusion, this paper has discussed the properties of optical
beams propagating in nonlocal nonlinear Kerr-like media. We have
shown that nonlocality leads to a variety of novel phenomena. 
In particular, it modifies the stability of plane waves, depending 
on whether the spectrum of the nonlocal response function is positive
definite or not.  
If the spectrum is positive definite then nonlocality always tends to
suppress MI, but can never eliminate it completely. 
If the spectrum is not positive definite then nonlocality may lead to the
appearance of higher order instability bands for focusing media. 
For defocusing media the effect of  nonlocality can be even more dramatic as it may actually initiate instability of the plane waves, 
leading to the formation of localised structures.

In local Kerr media multi-dimensional beams are unstable 
and collapse if they have a power above a certain critical value.
We have shown that a nonlocal nonlinearity always will arrest a 
collapse and allow for the formation of stable multi-dimensional
solitons. Furthermore, we have demonstrated
that quadratic solitons formed in the process of second harmonic
generation are equivalent to solitons of a nonlocal medium with
an exponential response function. Finally, we have shown  that
nonlocality induces attraction of normally repelling dark solitons
and allows for the formation of stable bound states.

In this review we have specifically concentrated on systems with spatially nonlocal,  Kerr  responses.  However, in many physical systems (BECs \cite{ParSalRea98} as well as pulse propagation in non-isotropic media), the Kerr response is composed of a 
nonlocal (delayed) response supplemented by a local contribution within a 
certain ratio. In the modelling of ultrashort pulse propagation nonlocality 
applies to the temporal response. In that case the causality principle requires the nonlocality to be represented by non-symmetric response functions, and hence the 
results presented in this review are not applicable to this situation. For 
more details on temporal aspects of nonlocality  see \cite{Couairon00,Berge00,Wyller01} and the references therein.


This research is supported by
the Danish Technical Research Council (Grant No.~26-00-0355) and  the
Australian Research Council. J.~Wyller acknowledges support from  The Research
Council of Norway under the grant No.~153405/432.


\end{document}